\documentclass{iau}
\usepackage{graphicx}

\usepackage{epsfig,subfigure,amsmath}
\usepackage{color}
\usepackage{hyperref}
\usepackage{natbib}

\definecolor{purple}{rgb}{1,0,1}

\newcommand{\lcdm}{$\Lambda$CDM}
\newcommand{\hmpc}{$h^{-1}$Mpc}
\newcommand{\gmpc}{$h^{-1}$Gpc}

\newcommand{\beq}{\begin{equation}}
\newcommand{\eeq}{\end{equation}}

\title[Distinguishing f(R) gravity with cosmic voids]{Distinguishing f(R) gravity with cosmic voids}

\author[P. Zivick and P.M. Sutter]   %% give here short author list %%
{P. Zivick$^{1}$\thanks{email: {\tt zivick.1@osu.edu}} and
P. M. Sutter$^{1,2,3}$}

\affiliation{
$^1$Center for Cosmology and AstroParticle Physics, Ohio State University, Columbus, USA \\
$^2$Sorbonne Universit\'{e}s, UPMC Univ Paris 06, UMR7095, F-75014, Paris, France \\
$^3$CNRS, UMR7095, Institut d'Astrophysique de Paris, F-75014, Paris, France \\
}

\pubyear{2014}
\volume{308}  %% insert here IAU Symposium No.
\pagerange{}
\jname{The Zeldovich Universe: Genesis and Growth of the Cosmic Web}
\editors{Rien van de Weygaert, Sergei Shandarin, Enn Saar, and Jan Einasto}

\begin{document}

\maketitle

\label{firstpage}

\begin{abstract}
We use properties of void populations identified in $N$-body
simulations to forecast the ability of upcoming galaxy surveys 
to differentiate models of f(R) gravity from \lcdm~cosmology. 
We analyze simulations 
designed to mimic the densities, volumes, and clustering statistics of 
upcoming surveys, using the public {\tt VIDE} toolkit. 
We examine void abundances as a basic probe at redshifts 1.0 and 0.4. We find that
stronger f(R) coupling strengths produce voids up to $\sim 20\%$ larger
in radius, leading to a significant shift in the void number
function. As an initial estimate of the constraining power of voids, 
we use this change in the number function to 
forecast a constraint on the coupling strength of $\Delta f_{R0} = 10^{-5}$.

\end{abstract}


% -----------------------------------------------------------------------------
% -----------------------------------------------------------------------------
\section{Introduction}

Modifications of gravity provide one way to explain the observed expansion of the universe. One such proposed theory is the \textit{f(R)} class of models, which contain relatively simple modifications to General Relativity (GR). This particular model incorporates the chameleon mechanism~\citep{Khoury:2004} that screens the fifth force in high density regions while leaving it unscreened in low density regions, strengthening the force of gravity.

Studying these underdense regions, called voids, could provide a way to test \textit{f(R)} gravity. Already voids have been used as a potential diagnostic for examining other models, such as coupled dark energy~\citep{Sutter2014d}. So far, current void-based studies of modified gravity
~\citep[e.g.,][]{Li:2012} have only focused on predictions for present-day conditions and ignored realistic survey effects. In this work, we mimic upcoming galaxy redshift surveys such as Euclid~\citep{Euclid:2011} 
and provide an initial estimate of the constraining power of void statistics.

% -----------------------------------------------------------------------------
% -----------------------------------------------------------------------------
\section{Analysis and Results}
\label{sec:approach}

We analyzed six simulation realizations from \citet{Zhao:2011a}. Three models with differing values for structure formation in the universe, expressed by $|f_{R,0}|$ with values $10^{-4}$ (F4), $10^{-5}$ (F5), and $10^{-6}$ (F6), were examined in addition to general relativity (GR). Each simulation box contained $1024^3$ dark matter particles and had a cubic volume of 1.5 \gmpc~per side. For analysis we selected snapshots at redshifts $z=0.43$ and $z=1.0$ and subsampled the DM particles to a mean density of $\bar{n} = 4 \times 10^{-3}$ per cubic \hmpc. This choice of redshift, density, and volume is designed to represent a 
typical space-based galaxy survey such as Euclid. Finally, we perturbed particle positions according to their peculiar velocities. We chose to ignore the effects of galaxy bias, as \citet{Sutter2013a} demonstrated that watershed void properties are relatively insensitive to bias.

Voids were identified using the publicly available Void Identification and Examination (VIDE) toolkit \citep{Sutter2014e}, which uses a substantially modified version of ZOBOV \citep{Neyrinck2008}. For this work, voids must be larger than the mean particle separation (in our case, $~1~$\hmpc) and have central densities higher than 0.2 of the mean particle density $\bar n$.
 
\begin{figure}
  \centering
{\includegraphics[type=eps,ext=.eps,read=.eps,width=0.49\columnwidth]{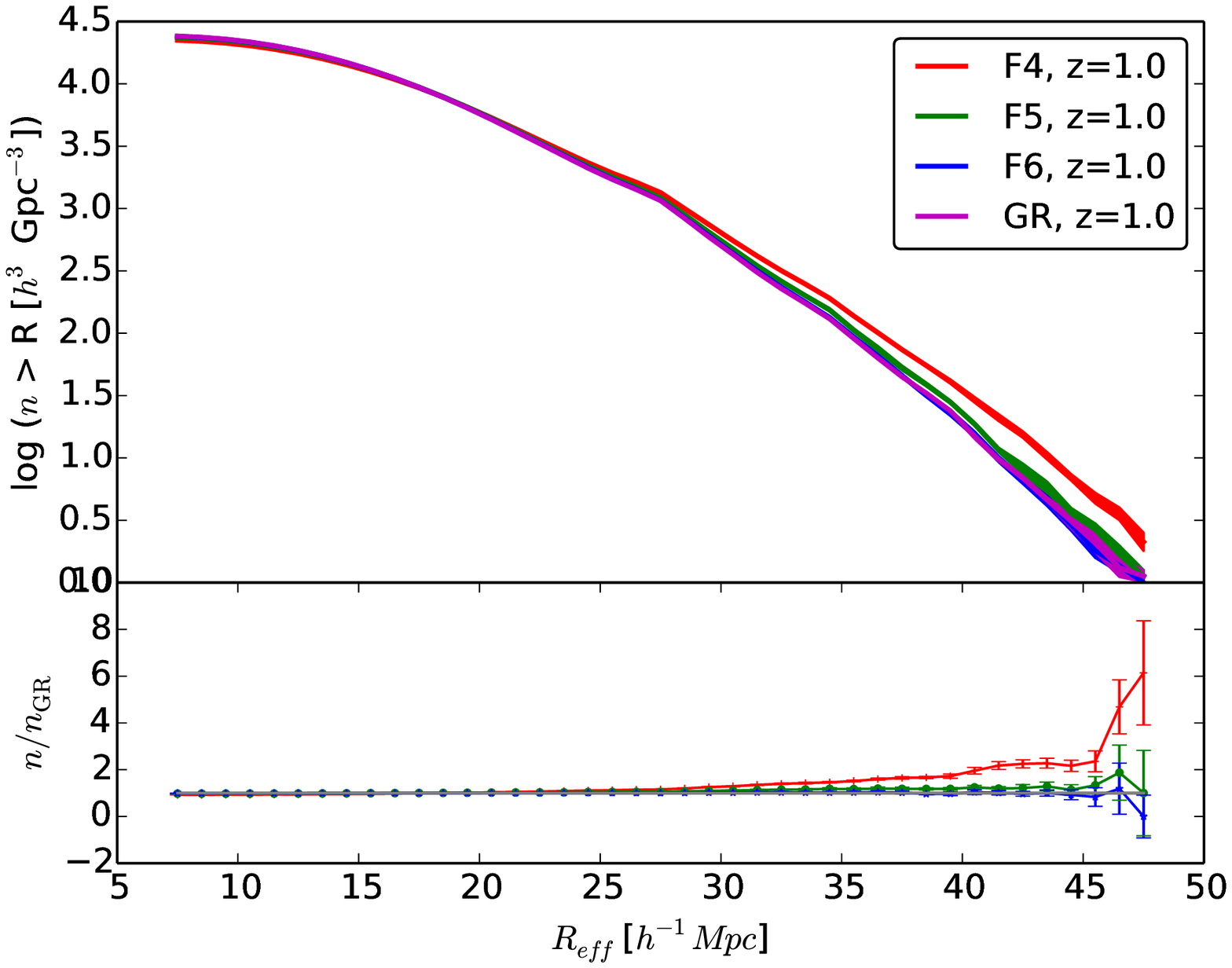}}
{\includegraphics[type=eps,ext=.eps,read=.eps,width=0.49\columnwidth]{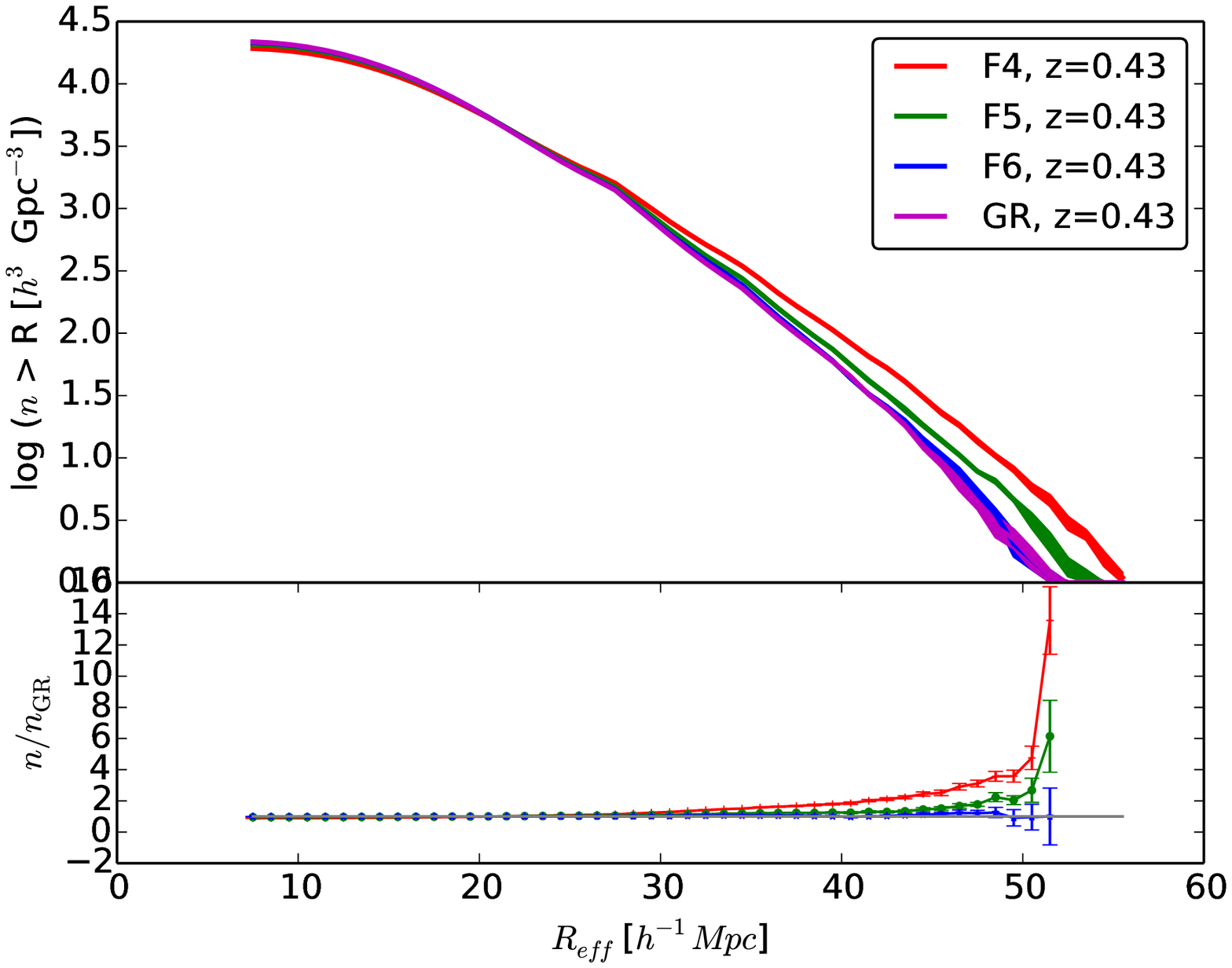}}
   \caption{Cumulative void number functions. Shown are the abundances (top) and relative abundances (bottom) for \lcdm~(purple) and modified gravity models F4 (red), F5 (green), and F6 (blue) from realistically subsampled dark matter particle simulations plotted against the effective void radius. The solid lines are the mean number functions of the six realizations, and the shaded regions are the 1$\sigma$ cosmic variances. Larger values of $|f_{R_{0}}|$ cause the fifth force to turn on at earlier ages, accelerating the evacuation of matter compared to \lcdm.
            }
\label{fig:numberfunc}
\end{figure}
 
 Figure~\ref{fig:numberfunc} shows the cumulative number function from \lcdm~ and \textit{f(R)} simulations at redshifts $z=1.0$ and $z=0.43$. We can see that F4 clearly contains larger voids than in the \lcdm~simulation at both redshifts. With weaker coupling strengths, one will notice that at high redshift, F5 and F6 are not able to separate from GR. At lower redshift, the F5 model becomes distinguishable at roughly the three sigma level from GR at radii as small as 35 \hmpc . Even the F6 model around 45 \hmpc~separates from GR, albeit by a relatively small amount. The gain in large voids is balanced by a loss of small voids, implying that the fifth force is accelerating the dissipation of interior void walls.

% -----------------------------------------------------------------------------
% -----------------------------------------------------------------------------
\section{Conclusions}
\label{sec:conclusions}

These features align with what one would reasonably expect to see from the $f(R)$ models. At higher redshift, the voids have not yet emptied out. Until the local densities pass a low enough threshold, the fifth force will remain screened, making the $f(R)$ models appear identical to GR. Simultaneously, the modified gravity 
mechanism only affects particle acceleration, and so the differences grow larger with time. Thus the strongest force, F4, produces the greatest number of large voids. 

An initial Fisher forecast places the constraint on measuring $|f_{R0}|$ at roughly $\Delta f_{R0} = 10^{-5}$, indicating that for stronger fifth forces, a detection may well be possible, especially at lower redshift where there is more statistical power.

% -----------------------------------------------------------------------------
% -----------------------------------------------------------------------------

\footnotesize{
  \bibliographystyle{mn2e}
  \bibliography{Zivick2014}
}

\end{document}